# EAR-U-Net: EfficientNet and attention-based residual U-Net for automatic liver segmentation in CT

Jinke Wang[1, 2, *], Xiangyang Zhang[1], Peiqing Lv[1], Lubiao Zhou[1], Haiying Wang[1]

[1]*School of Automation, Harbin University of Science and Technology, Harbin, 150080, China*

[2]*Rongcheng College, Harbin University of Science and Technology, Rongcheng, 264300, China*

**Abstract**：

    ***Purpose***: This paper proposes a new network framework called EAR-U-Net, which leverages EfficientNetB4, attention gate, and residual learning techniques to achieve automatic and accurate liver segmentation.

    ***Methods***: The proposed method is based on the U-Net framework. First, we use EfficientNetB4 as the encoder to extract more feature information during the encoding stage. Then, an attention gate is introduced in the skip connection to eliminate irrelevant regions and highlight features of a specific segmentation task. Finally, to alleviate the problem of gradient vanishment, we replace the traditional convolution of the decoder with a residual block to improve the segmentation accuracy.

    ***Results***: We verified the proposed method on the LiTS17 and SLiver07 datasets and compared it with classical networks such as FCN, U-Net, Attention U-Net, and Attention Res-U-Net. In the Sliver07 evaluation, the proposed method achieved the best segmentation performance on all five standard metrics. Meanwhile, in the LiTS17 assessment, the best performance is obtained except for a slight inferior on RVD. Moreover, we also participated in the MICCIA-LiTS17 challenge, and the Dice per case score was 0.952.

    ***Conclusion***: The proposed method's qualitative and quantitative results demonstrated its applicability in liver segmentation and proved its good prospect in computer-assisted liver segmentation.

**Keywords**: Liver segmentation, EfficientNet, U-Net, Residual, Attention

## 1. Introduction

    According to Cancer Analysis 2020 [1], the malignant liver tumor is the sixth most common cancer and the second leading cause of cancer deaths. To help the physicians make accurate assessment and treatment at an early stage, the computed tomography (CT)-based segmentation is widely used in the screening, diagnosis, and tumor measurement. However, liver and liver tumors show a high degree of variability in shape, appearance, and location and vary from person to person (as shown in Fig. 1),

resulting in the manual segmentation of the liver being labor-intensive and error-prone. Therefore, how to segment the liver automatically and accurately has become a challenging and valuable task.

In recent years, many automatic liver segmentation approaches have emerged because of their ability to eliminate subjective factors and improve the accuracy and efficiency of diagnosis. These methods can be divided into two categories: (1) handcraft feature-based methods and (2) deep learning-based methods.

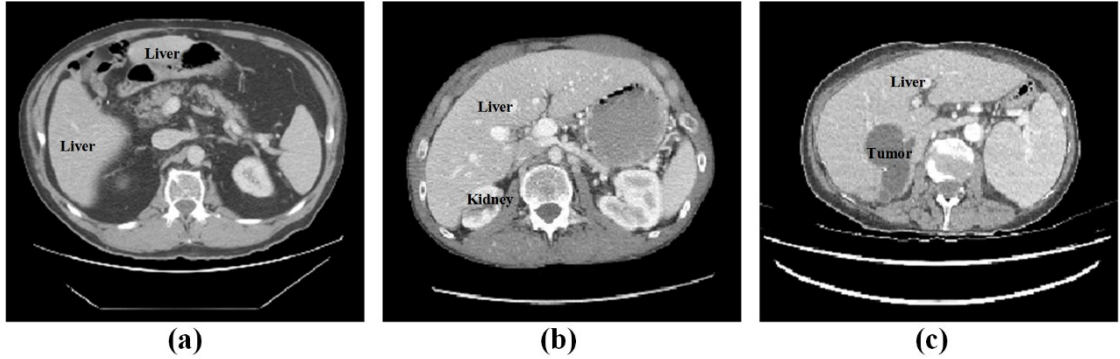

**Fig. 1.** Liver CT with significant variations. (a) liver consists of discontinuous regions (b) liver with an adjacent organ of low contrast (c) liver with the tumor

The handcraft feature-based methods mainly include region growth [2-3], thresholding [4-5], model-based methods [6-8], and machine learning-based methods [9-13]. These methods manually extract features from the input image, such as intensity, shape, edge, texture, or some transformation coefficients, and then generate the contour or region of the liver according to the local feature differences. Le et al. [14] proposed a 3D fast marching algorithm and single hidden layer feedforward neural network. First, the 3D fast marching algorithm is used to create the initial marker region. Then the single hidden layer feedforward neural network (SLFN) is employed to classify the unlabeled voxels, and finally, the liver tumor boundary was extracted and refined by post-processing. Singh et al.'s improved k-means clustering method [15] refines the clustering through ant colony optimization. Their accuracy and segmentation time of liver segmentation is superior to those of previous technologies. Although these methods achieved good accuracy in limited sample space, most are semi-automatic approaches with poor stability, require artificial feature engineering, and have limited representation capabilities.

Deep learning-based methods have been popular in the computer vision community in recent years. Specifically, CNN has developed rapidly from classification network AIexNet [16] to ResNet [17]. However, unlike classification tasks, liver segmentation is pixel-driven classification, which makes the segmentation task more complicated than classification. The most popular deep learning-based segmentation methods include full convolutional neural network (FCN) [18], U-Net [19] and its variants [20-23], and auto encoder-decoder neural networks (AED) [24].

Long et al. [18] suggested the novel FCN by replacing the fully connected layer with a convolutional layer and restoring the image through de-convolution. Their pixel-level prediction is then widely used in semantic segmentation for its end-to-end framework. Ben-Cohen et al. [25] employed FCN for liver segmentation and lesions detection for the first time. Sun et al. [26] designed a multi-channel FCN to segment liver tumors from multi-phase contrast-enhanced CT (CECT) images. In the high-level layer after feature extraction, feature fusion is performed on multi-phase CECT to improve the segmentation accuracy. Zhang et al. [27] designed a cascaded FCN for rough segmentation of the liver. For post-processing, they used different classic segmentation models, such as level set, graph cut, and the conditional random field (CRF). Such a segmentation approach that combines deep learning with machine learning has been effectively applied in many fields.

Based on FCN, Ronneberger et al. [19] proposed U-Net in the same year. Compared with FCN, U-Net designed an elaborate skip connection, perfect decoding structure, and higher segmentation accuracy. Jin et al. [28] proposed a hybrid deep attention-aware network (RA-U-Net) to extract liver and tumor. It is the first work that employs a residual attention mechanism to process medical volumetric images. Wardhana et al. [29] proposed a 2.5D model to segment liver and tumor. This model allows the network to equip a deeper and wider network while containing 3D information. Further, Li et al. [30] propose a novel hybrid densely connected U-Net (H-DenseUNet), which combines 2D and 3D networks to fully integrate the information within and between the slices to achieve higher segmentation accuracy.

The automatic encoder-decoder neural network has also received significant attention in the field of liver segmentation. Lei et al. [31] propose a deformable encoder-decoder network (DefED-Net) for liver and liver tumor segmentation. First, they used deformable convolution to enhance the feature representation ability of the DefED network. Then they designed a trapezoidal atrous pyramid pool (ASPP) module based on a multi-scale expansion rate and achieved a Dice of 0.963 on the LiTS17-training dataset. Tummala et al. [32] developed a multi-scale residual dilated encoder-decoder network to segment liver tumors. First, the proposed network segments the liver and then extracts tumors from the liver ROIs. Next, they reduce the image to different resolutions at each scale and apply regular convolution, dilation, and residual connections to capture a wide range of conceptual information.

However, most deep learning-based networks are not sensitive to the details of liver images, and the feature results obtained by de-convolution are relatively smooth. Although the U-Net model can enhance the decoder's feature learning through skip connections and performs well in medical image segmentation, U-Net's segmentation of image details is still not satisfactory. Besides, the number of layers and parameters is small. Therefore, it is easy to result in over-fitting problems. Moreover, U-Net uses a pooling layer in the process of down-sampling, which may lose many image features. In addition, the learned shallow information is limited, and it is prone to result in over-/under segmentation error after connecting with the in-depth information. Finally, as the depth of the network increases, the problem of gradients vanishing may occur. Also, most automatic encoding and decoding neural networks are variants of FCN and U-Net, which could have similar disadvantages.

To alleviate the problems mentioned above, this paper proposes a novel end-to-end U-Net-based framework, called EAR-U-Net[1], leveraging EfficientNetB4, attention

---

[1] Our code is publicly available at https://github.com/ZhangXY-123/Model/blob/master/EAR_Unet.py

gate, and residual learning techniques for automatic and accurate liver segmentation.

The main contributions of this paper are as follows:
- Use a modified EfficientNet-B4 as the encoder to extract more feature information in the encoder stage.
- Add an attention gate to the original skip connection to eliminate irrelevant regions and focus on the liver area to be segmented.
- Employ the residual structure to replace the convolutional layer in the U-Net decoder and add a batch normalization layer to eliminate the gradient vanishment problem, accelerate the convergence speed, and achieve higher accuracy.

The structure of the whole paper is as follows: the second section introduces the related work, and in the third section, we descript the proposed EAR-U-Net framework in detail. The fourth section provides the experimental results and discussion, and in the final fifth section, we summarize the whole work and give a future outlook.

## 2 Related works

This section introduces the related work, including EfficientNet, attention mechanism, and residual learning.

### 2.1 EfficientNet

EfficientNet has attracted extensive attention because it can balance the model's depth, width, and image resolution. Previously, in the process of deep learning model training, the most commonly used method to improve the model accuracy was to expand the width of the network, increase the depth of the network and enhance the resolution of the input image. For example, VGGNet-11 [33] is extended to VGGNet-19, expanding the depth of the network. GoogLeNet [34] proposed the inception module to increase the network depth and width. However, the balance of network width, depth, and resolution is still not fully considered. Thus, Tan and Le [35] proposed EfficientNet, which designed a new model scaling method to balance the model's depth, width, and resolution through composite coefficients. Table 1 lists the structures of the eight models (EfficientNetB0-EfficientNetB7). This paper adopts the improved EfficientNetB4.

**Table 1** EfficientNetB0-EfficientNetB7

| Medol | Inputsize | Width coefficient | Depth coefficient |
|---|---|---|---|
| EfficientNetB0 | 224×224 | 1.0 | 1.0 |
| EfficientNetB1 | 240×240 | 1.0 | 1.1 |
| EfficientNetB2 | 260×260 | 1.1 | 1.2 |
| EfficientNetB3 | 300×300 | 1.2 | 1.4 |
| EfficientNetB4 | 380×380 | 1.4 | 1.8 |
| EfficientNetB5 | 456×456 | 1.6 | 2.2 |
| EfficientNetB6 | 528×528 | 1.8 | 2.6 |
| EfficientNetB7 | 600×600 | 2.0 | 3.1 |

EfficientNet has been widely used in image classification and segmentation. For example, Chetoui et al. [36] used EfficientNet to achieve the most advanced performance in Diabetic retinopathy (DR) work. Kamble et al. [37] employed the EfficientNet as an encoder, combined with U-Net++, and achieved high accuracy in optic disc (OD) segmentation. Messaoudi et al. [38] used EfficientNet to convert a 2D classification network into a 3D semantic segmentation of brain tumors, which also obtained satisfactory performance.

**2.2 Attention mechanism**

The attention mechanism has been popular in classification and segmentation communities because of its lower complexity and fewer parameters than CNN and RNN and its ability to capture global and local information [39-41]. The attention mechanism in biological perception is mainly used to select a subset of perception information for complex processing and to perform prohibition operations on all organ inputs. The basic idea of the attention mechanism is to allow the system to learn attention, ignore irrelevant information, and focus on useful essential information.

The attention mechanism can be classified into hard and soft attention mechanisms [42]. The hard attention mechanism needs to predict the attention region, which is usually trained by reinforcement learning. The soft attention mechanism has been widely used in computer vision by selectively ignoring part of the information to re-weight and aggregate the rest. Oktay et al. [43] designed the attention gate to suppress irrelevant information in skip connections. Attention gate improves the prediction ability of U-Net without reducing the computational efficiency. Fu et al. [44] proposed a dual attention mechanism network with channel attention and location attention mechanisms, enhancing the dependence between different channels and locations to improve the model's accuracy. Sinha and Dolz et al. [45] proposed the multi-scale self-guided attention, which obtains global features through multi-scale strategy, and then introduces the learned global features into the attention module. This method has been verified in the segmentation experiments of abdominal organs, cardiovascular structures, and brain tumors. Liu et al. [46] proposed a cascaded atrous dual attention U-Net for tumor segmentation. The proposed network structure connects the features of 3D liver segmentation with those of 2D tumor segmentation and then embeds double attention gates in the skip structure of the 2D model. They evaluated the proposed method on databases of different organs and confirmed good performance.

**2.3 Residual learning**

The residual structure has attracted extensive attention because it solves the problems of gradient vanishing and explosion. He et al. [17] proposed the residual module for the first time in 2015. Before that, constructing a deep network usually involves extracting more information, but it also brings many problems. The most serious difficulty is that the gradient will disappear or explode. The traditional solution employs gradient clipping and weight regularization, but this will still cause network degradation. Furthermore, as the number of network layers increases, the training accuracy will tend to be saturated. Likewise, if the number of layers continues to grow, the training accuracy will decline. However, the residual structure can solve the problems of gradient vanishment/explosion and alleviate network degradation.

Employing the residual structure in the segmentation task can often obtain an improvement in accuracy. Mourya et al. [47] designed a dilated deep residual network (DDRN) for liver segmentation, cascading a combination of three parallel DDRNs with a fourth DDRN to obtain the final result and achieve excellent segmentation results. Yu et al. [48] introduced the residual structure into the 3D U-Net and constructed a new 3D residual U-Net framework. Compared with the previous method, this method is more accurate and stable for extracting hepatic blood vessels. Alom et al. [49] proposed a recurrent residual CNN (RRCNN) based on the U-Net model, which uses the recursive residual convolutional layer for feature accumulation to represent segmentation tasks better. This method has been verified in multiple medical image segmentation datasets. The emergence of residual modules brings more space for the improvement of network depth.

## 3 Method

This section introduces the architecture of the proposed EAR-U-Net in detail.

### 3.1 Model Architecture

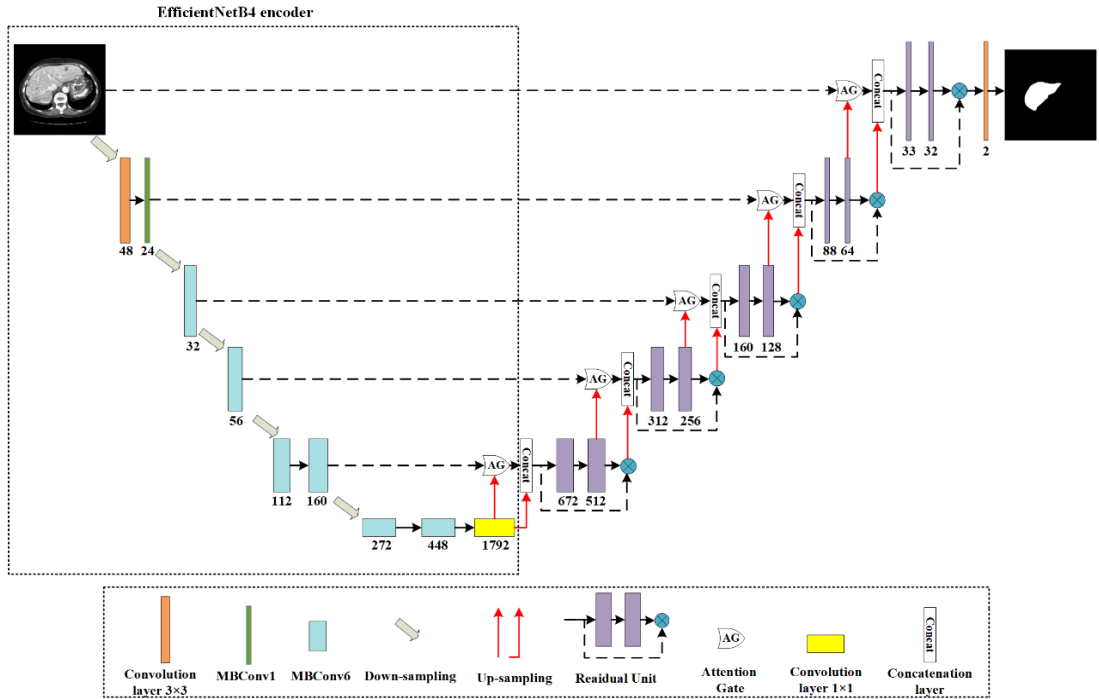

**Fig. 2.** The architecture of the proposed EAR-U-Net

The proposed network EAR-U-Net consists of an encoder and decoder (Fig. 2). Considering the limitation of computing resources, we employ the modified EfficientNetB4 as the encoder. The encoder consists of nine stages (Table 2), including a 3×3 convolutional layer, 32 mobile reversed bottleneck convolutional (MBConv) structures, and a 1×1 convolutional layer. The decoder is composed of five up-sampling and a series of convolution operations. The features extracted by the encoder are restored to the original image size, and then the segmentation results are obtained. To reduce the noise response and focus on specific features, we add an attention gate to the skip connection to make the segmented liver more accurate. The addition of the residual structure can increase the depth of the network. In the residual block, batch normalization (BN) and ReLU activation are performed after each convolution. The introduction of batch normalization can eliminate gradient diffusion and vanishment and accelerate the convergence of the network. Then use ReLU to perform non-linear processing to improve the non-linear expression ability y of the network.

Table 2 The structure of the encoder

| Stage | Operator | Resolution | Channels | Layers |
|---|---|---|---|---|
| 1 | Conv3×3 | 256×256 | 48 | 1 |
| 2 | MBConv1,k3×3 | 128×128 | 24 | 2 |
| 3 | MBConv6,k3×3 | 128×128 | 32 | 4 |
| 4 | MBConv6,k5×5 | 64×64 | 56 | 4 |
| 5 | MBConv6,k3×3 | 32×32 | 112 | 6 |
| 6 | MBConv6,k5×5 | 16×16 | 160 | 6 |
| 7 | MBConv6,k5×5 | 16×16 | 272 | 8 |
| 8 | MBConv6,k3×3 | 8×8 | 448 | 2 |
| 9 | Conv1×1 | 8×8 | 1792 | 1 |

The MBConv structure comprises a 1×1 convolution, a Depthwise convolution, a sequence-and-exception (SE) module, a 1×1 convolution for dimension reduction, and the dropout layer (Fig. 3). After the first 1×1 convolution and Depthwise Conv convolution, BN and Swish activation operations are conducted, and the second 1×1 convolution only performs BN operations. To fuse more feature information, we add a shortcut connection. The shortcut connection only exists when the shape of the feature matrix of the input MBConv structure is the same as that of the output feature matrix.

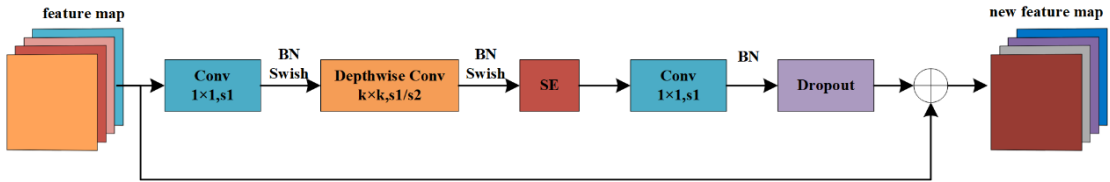

**Fig. 3**. MBConv Block

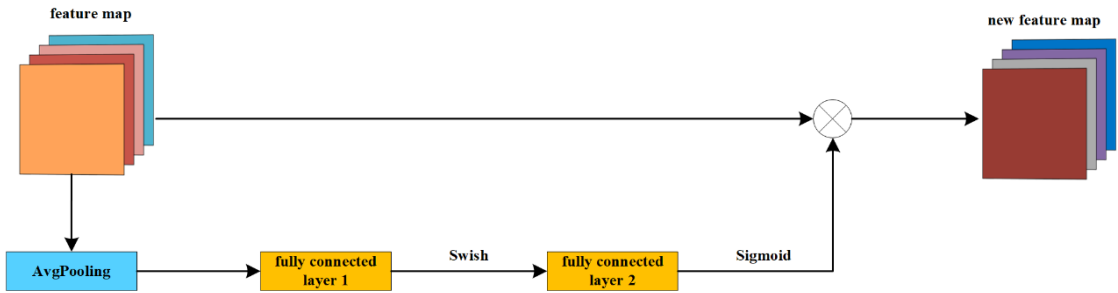

**Fig. 4.** Squeeze and Excitation Block

The SE module has dramatically improved the accuracy in image classification,

target detection, and image segmentation. The SE module used in this paper (Fig. 4) consists of a global average pooling, two fully connected layers, and a Sigmoid activation function. In addition, the Swish activation function is added between the two full connection layers. Assuming input an image H×W×C, first, stretch it into 1×1×C through the global pooling and fully connected layers, and then multiply it with the original image to give weight to each channel. In this way, the SE module enables the network to learn more liver-related feature information.

Attention gate is a kind of attention mechanism that could automatically focus on the target area, suppress the response of irrelevant regions, and highlight the feature information crucial to a specific task, whose structure is shown in Fig. 5. First, *g* and *x* go through the 1×1 Conv operation in parallel, and implement the add operation at the corresponding points. Then perform the ReLU activation, 1×1 Conv and Sigmoid function operations sequentially, and resample to get the attention coefficient α. Finally, the attention coefficient α is multiplied by the input coding matrix *x* to obtain the final output.

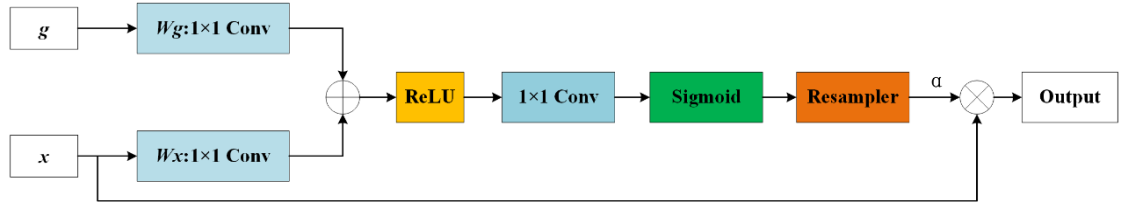

**Fig. 5**. Schematic of the attention gate (*g* is the decoding matrix, and *x* is the encoding matrix)

## 4 Experiments

This section first describes the datasets used in the paper, the image pre-processing, the dataset augmentation, and the implementation details. Then we provide the loss function and evaluation metrics of the evaluation. Finally, the experimental results are shown and analyzed, and the method's limitation is discussed as well.

### 4.1 Experimental setup

#### 4.1.1 Image dataset

In this experiment, we used the labeled training sets of the LiTS17 and SLiver07 datasets for testing. The LiTS17-training dataset consists of 131 abdominal CT scans, with a large varying in-plane resolution from 0.55 mm to 1.0 mm and the inter-slice spacing from 0.45 mm to 6.0 mm. The number of slices ranges from 75 to 987. The size of each slice is 512 × 512. The SLiver07 training dataset consists of 20 CT scans, with in-plane resolution from 0.55 mm to 0.8 mm and inter-slice spacing from 1.0 mm to 3.0 mm. The number of slices ranges from 64 to 394, and each slice's size is 512 × 512 (shown in Table 3).

**Table 3** Specifications of LiTS17 and SLiver07 datasets

| Datasets | Training data | Inter-pixel spacing | Inter-slice spacing | Number of slices | Resolution |
| --- | --- | --- | --- | --- | --- |
| LiTS17 | 131 | 0.55mm-1.0mm | 0.45mm-6.0mm | 75-987 | 512×512 |
| Sliver07 | 20 | 0.55mm-0.8mm | 1.0mm-3.0mm | 64-394 | 512×512 |

### 4.1.2 Image preprocessing

We first set the Hounsfield intensity to [-200, 200] to exclude irrelevant details and employ histogram equalization to enhance the contrast of the image. Then the CT image is down-sampled and resampled on the cross-section. Next, the spacings of the z-axis of all scans are adjusted to 1mm to make the data more balanced. After that, we locate the slices with the liver and expand 20 slices outward to the edge slices at both ends. Finally, to save training time and reduce the memory requirements, we set each image's size to 256×256. The whole workflow is depicted in Fig. 6.

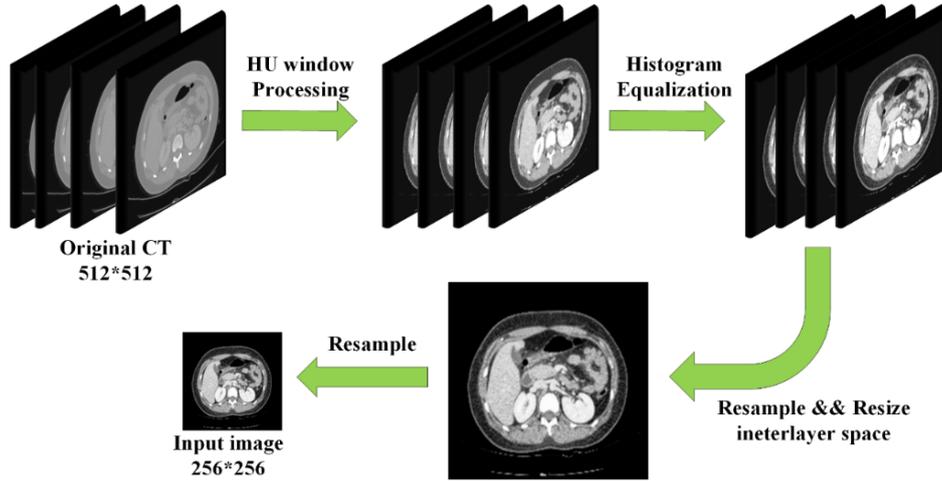

**Fig. 6.** Flowchart of image pre-processing.

### 4.1.3 Dataset augmentation

Considering the SLiver07-training dataset has a small amount of data, we enhanced the image data to improve the model's generalization ability and prevent the overfitting problem. Meanwhile, we zoom the data with mirror flip, rigid and elastic deformations. Fig. 7 illustrates some cases using different enhancement strategies.

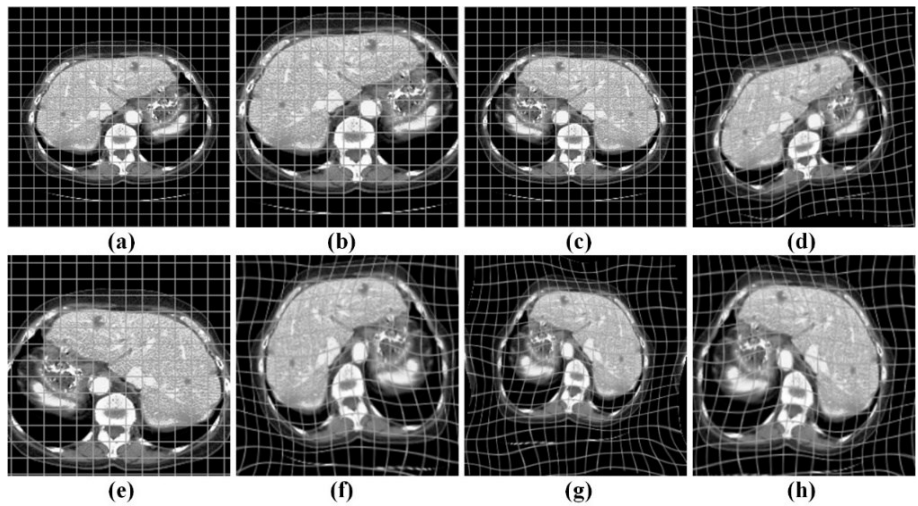

**Fig. 7.** Data augmentation. (a) original CT with gird (b) zoom (c) mirror flip (d) elastic deformation (e) zoom and mirror flip (f) zoom and elastic deformation (g) mirror flip and elastic deformation (h) zoom, mirror flip and elastic deformation

### 4.1.4 Implementation details

We run all the experiments on a workstation with Ubuntu 18.04 operating system, graphics card RTX2080Ti, RAM 32G, single CPU Intel Xeon Silver 4110, and using the Pytorch1.8 deep learning framework for implementation. In the network training, we set the batch size to 16, set the epoch to 60, choose Adam as the optimizer, and set the learning rate to 0.001 (Table 4).

**Table 4** Training parameters

| Training parameter | Batch size | epoch | optimizer | Learn rate |
|---|---|---|---|---|
| value | 16 | 60 | Adam | 0.001 |

### 4.2 Loss function definition

The loss function makes an essential impact on the performance of CNN. In medical image segmentation, since ROI only covers a small area, and thus it is prone to lead to a sharp decline of the loss function to the local minimum during training, which may result in a significant segmentation deviation. However, Cross-entropy [50] is able to measure the difference between two different probability distributions in the same random variable. The smaller the value of cross-entropy, the more accurate the prediction of the model. Therefore, Cross-entropy can achieve good results in the segmentation network of pixel-level classification. The Binary Cross-entropy is defined below.

$$L_{BCE}(y, \hat{y}) = -(y \log(\hat{y}) + (1-y) \log(1-\hat{y})) \quad (1)$$

where $y$ represents the actual value and $\hat{y}$ represents the predicted result. Dice coefficient is one of the standard metrics to evaluate the segmentation effect. It can also be used to measure the distance between the segmentation result and the label [51]. As a loss function, Dice Loss (DL) performs well in processing unbalanced datasets and can effectively reduce segmentation deviation caused by unbalanced ROI area and background. The DL used in this paper is defined in Eq. (2).

$$DL(y, \hat{p}) = 1 - \frac{2y\hat{p}+1}{y+\hat{p}+1} \quad (2)$$

where value 1 is added in numerator and denominator to ensure that the function is not undefined in edge case scenarios such as when $y = \hat{p} = 0$.

### 4.3 Evaluation Metrics

In this paper, we choose five commonly used metrics for evaluation, including Dice, volume overlap error (VOE), relative volume error (RVD), average symmetrical surface distance (ASSD), and Maximum Surface Distance (MSD). For Dice and VOE, the larger the value, the better the segmentation result, while ASSD, RVD and MSD are the opposite.

Assuming that $A$ is the segmentation result area of the liver, and $B$ is the ground truth, the five metrics can be defined as follows:
(1) Dice: The similarity of the two sets. The larger the value, the better the segmentation effect.

$$Dice(A, B) = \frac{2|A \cap B|}{|A|+|B|} \quad (3)$$

(2) VOE: The error between the predicted segmentation volume and the ground truth.

$$VOE(A,B) = 1 - \frac{|A \cap B|}{|A \cup B|} \quad (4)$$

(3) RVD: Used to determine whether the segmentation result is in an over- or under-segmentation.

$$RVD(A,B) = \frac{|B|-|A|}{|A|} \quad (5)$$

(4) ASSD: The average distance between the surfaces of segmentation results *A* and *B*, where *d*(*v*, *S*(*X*)) represents the shortest Euler distance from voxel *v* to the surface voxel.

$$ASSD(A,B) = \frac{1}{|S(A)|+|S(B)|}\left(\sum_{p \in S(A)} d(p, S(B)) + \sum_{q \in S(B)} d(q, S(A))\right) \quad (6)$$

(5) MSD: The max distance between the surfaces of segmentation results *A* and *B*, where *d*(*v*, *S*(*X*)) represents the shortest Euler distance from voxel *v* to the surface voxel.

$$MSD(A,B) = \max\{\max_{p \in S(A)} d(p, S(B)), \max_{q \in S(B)} d(q, S(A))\} \quad (7)$$

**4.4 Test on LiTS17-Training dataset**

In this section, we conducted experiments on the LiTS17-Training dataset. We randomly selected 121 sets of sans as the training and validation sets, while the remaining ten sets as the test set. To verify EAR-U-Net's performance, we first used the most commonly used DL as the loss function. Next, we performed comparative experiments and ablation experiments, respectively. Finally, to evaluate the effectiveness of DL + Binary Cross-Entropy Loss (BL), we select the combination of DL: BL = 1:1 as the loss function and take the classical models FCN [18], U-Net [19], attention U-Net [43], attention Res-U-Net and EAR-U-Net for comparison.

**4.4.1 Comparison with classical methods**

First, we use DL as the loss function and compare the classic network FCN[2], U-Net[3], Attention U-Net[4], and Attention Res-U-Net[5]. From Table 5, we can see that FCN results in the worst performance on Dice and VOE compared to the other four networks. On the other hand, compared with FCN, U-Net, Attention U-Net, and Attention Res-U-Net, the proposed EAR-U-Net model achieved the best performances on the four metrics (Dice, VOE, ASSD, and MSD) except for RVD. Specifically, its superiority on MSD is the most significant.

Therefore, EAR-U-Net enabled an improvement in the accuracy and stability of the segmentation. Besides, in terms of training time, EAR-U-Net is far less than U-Net, Attention U-Net, and Attention Res-U-Net, only more than FCN. However, in terms of test time, the EAR-U-Net is higher than other networks.

---

[2] The code is available at https://github.com/shelhamer/fcn.berkeleyvision.org
[3] The code is available at https://github.com/JavisPeng/u_net_liver/blob/master/unet.py
[4] The code is available at https://github.com/Andy-zhujunwen/UNET-ZOO/blob/master/attention_unet.py
[5] The code is available at https://github.com/ZhangXY-123/Model/blob/master/Res_Att_Unet.py

Table 5 Quantitative results among the five methods on 10 LiTS17-Training datasets

| Method | Dice (%) | VOE (%) | RVD (%) | ASSD (mm) | MSD (mm) | Training time | Testing time |
|---|---|---|---|---|---|---|---|
| FCN | 92.46±3.52* | 13.83±5.83 | -1.65±8.74 | 2.86±1.24 | 81.94±28.95 | **4h31m13s** | **33s** |
| U-Net | 94.08±2.06* | 11.12±3.65 | -0.48±5.58 | 3.07±2.08 | 66.03±27.91 | 7h49m13s | 36.4s |
| Attention U-Net | 94.37±2.27* | 10.58±4.04 | **0.37±6.91** | 2.91±1.57 | 82.03±31.43 | 8h56m35s | 36.8s |
| Attention Res-U-Net | 94.93±1.63* | 9.61±2.97 | 2.23±4.12 | 2.77±1.69 | 62.69±19.71 | 9h48m56s | 37.1s |
| EAR-U-Net | **95.95±0.76** | **7.77±1.42** | 0.50±2.36 | **1.29±0.35** | **35.96±20.62** | 6h45m54s | 41.2s |

Results are represented as mean and standard deviation. Note: ∗ indicates a statistically significant difference between the marked result and the corresponding one of our method at a significance level of 0.05.

To demonstrate the robustness of the proposed EAR-U-Net more intuitively, we depict the boxplot on the five metrics. From Fig. 8, we can see that the proposed EAR-U-Net exhibits strong stability on all five metrics. Specifically, for Dice (Fig. 8 (a)), the median of EAR-U-Net achieved the highest without outlier compared with the other four networks.

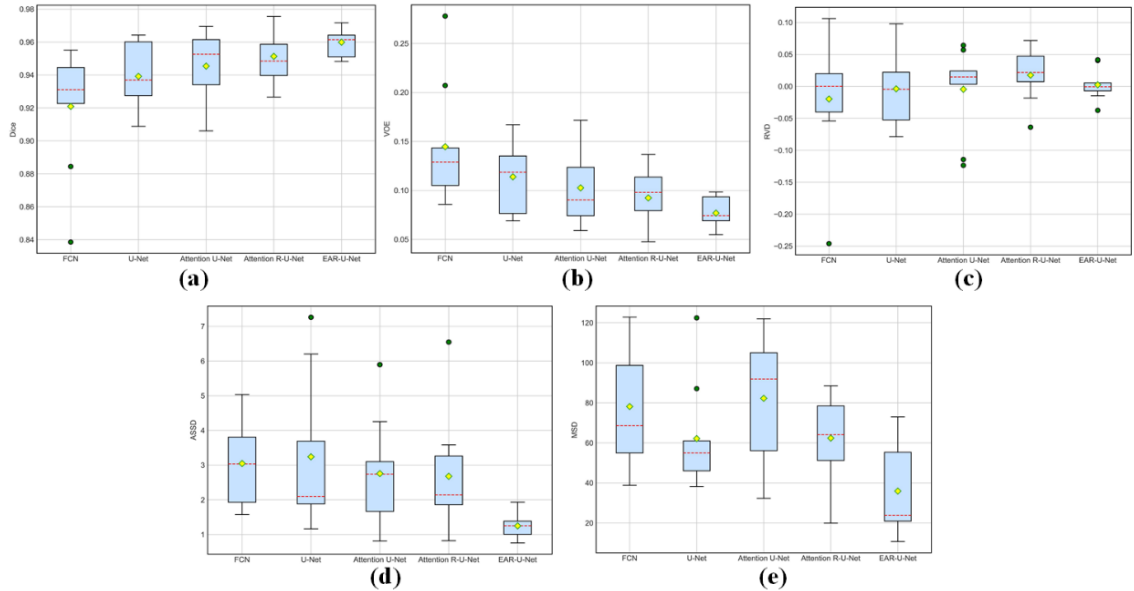

**Fig. 8.** Comparative analysis on five metrics. (a) Dice (b)VOE (c) RVD (d) ASSD (e) MSD

For VOE (Fig. 8 (b)), the median of EAR-U-Net is the lowest, with the highest stability. Besides, the median on RVD (Fig. 8 (c)) is closer to 0, but there are two outliers. Moreover, it shows extreme stability on ASSD (Fig. 8 (d)), and the median of MSD (Fig. 8 (e)) is far less than that of the other four networks.

Fig. 9 shows the loss curves of training and testing. From the figures, we can see that the loss value of the EAR-U-Net network is smoother and converges faster than other models.

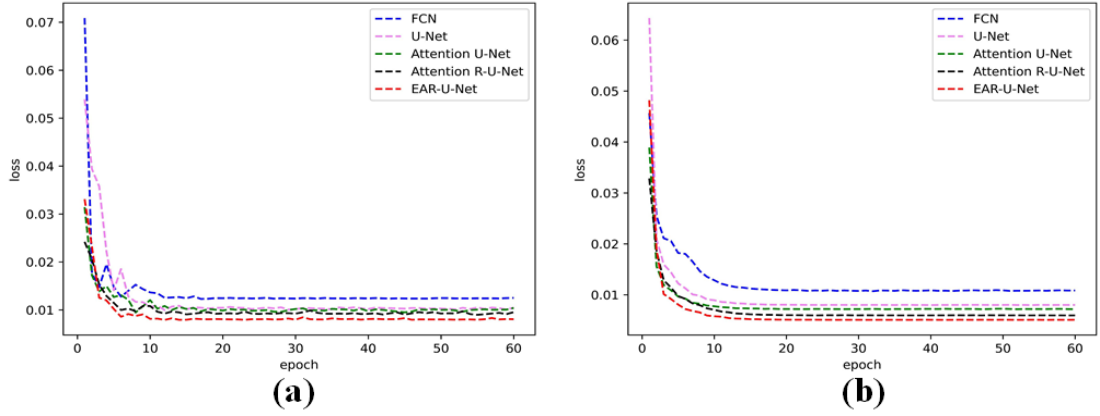

**Fig. 9.** Loss curves of different models on LiTS17 datasets (a) the training set (b) the validation set.

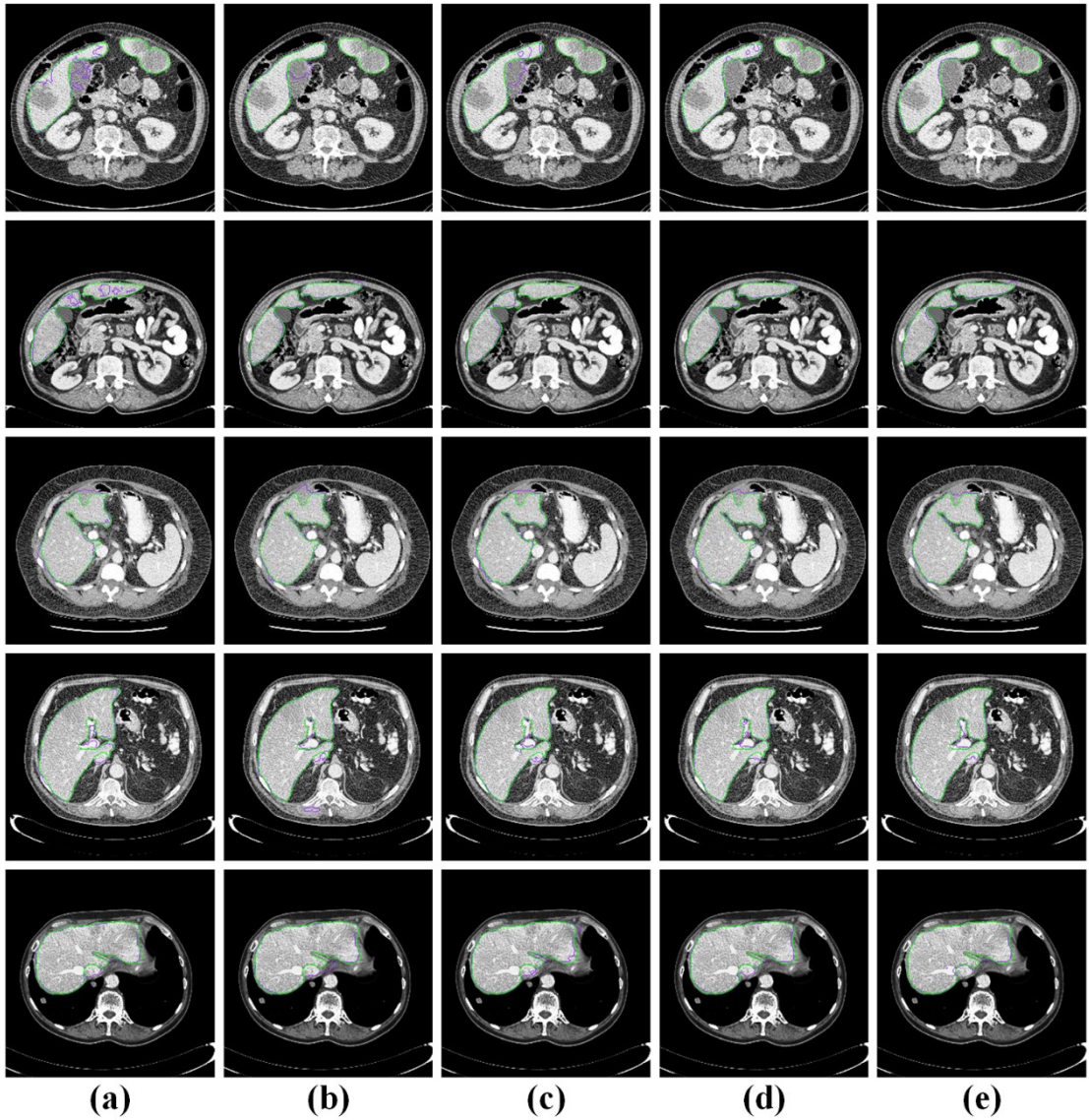

**Fig. 10.** Visualization of challenging cases. (a) FCN (b) U-Net (c) Attention U-Net (d) Attention Res-U-Net (e) EAR-U-Net. (The green line represents the ground truth, and the purple line represents the segmentation result of the corresponding method)

Fig. 10 shows some visualizations of challenging cases. The first and the second row are discontinuous liver regions. (i) In the first row, FCN, U-Net, and the Attention U-Net incorrectly segmented the gallbladder adjacent to the liver. Meanwhile, the Attention Res-U-Net showed a little under-segmentation error. On the contrary, the proposed EAR-U-Net segmented the liver almost perfectly. (ii) In the second row, FCN showed obvious over-segmentation error, while other models performed well. (iii) The third row illustrates the segmentation of the liver with interlobar fissure. FCN and U-Net showed under-segmentation errors, but U-Net, Attention Res-U-Net, and our proposed methods showed slight errors. (iv) The fourth row provided the liver area containing the portal vein. Again, we can see that FCN, U-Net, and Attention U-Net have mistakenly under-segmented the portal artery. Nevertheless, the effect of attention Res-U-Net and our model is much superior to the other three models. (v) The fifth row shows the liver region containing the inferior vena cava. It can be seen that, except for the complete liver segmentation by the proposed network, the other four networks all mistakenly segment the inferior vena cava as the liver. The above demonstrates that our proposed network has advantages in the discontinuous liver region, the liver region with adjacent organs, and portal veins.

**4.4.2 Ablation analysis on LiTS17-Training datasets**

To verify the optimality of the proposed network, we performed four comparative ablation experiments based on the Efficient module (E-U-Net), Efficient residual structures (ER-U-Net), and Efficient attention gate (EA-U-Net). Specifically, we use the DL loss function for training, with the test results shown in Table 6.

**Table 6** Quantitative analysis results of ablation experiments

| Method | Dice(%) | VOE(%) | RVD(%) | ASSD(mm) | MSD(mm) | Training time | Testing time |
|---|---|---|---|---|---|---|---|
| E-U-Net | 95.23±1.44* | 9.07±2.62 | **0.10±3.3** | 2.14±1.21 | 80.34±23.51 | **5h48m45s** | **40.2s** |
| EA-U-Net | 95.28±1.37* | 8.99±2.49 | 0.56±2.94 | 2.11±1.07 | 75.46±21.71 | 6h27m27s | 40.9s |
| ER-U-Net | 95.62±1.17* | 8.37±2.15 | 0.78±2.66 | 1.64±0.49 | 68.41±23.79 | 6h4s40s | 40.6s |
| EAR-U-Net | **95.95±0.76** | **7.77±1.42** | 0.50±2.36 | **1.29±0.35** | **35.96±20.62** | 6h45m54s | 41.2s |

Results are represented as mean and standard deviation. Note: ∗ indicates a statistically significant difference between the marked result and the corresponding one of our method at a significance level of 0.05.

Table 6 shows that EAR-U-Net has achieved the best results on the five standard metrics except for RVD. The employment of residual structures enables a significant improvement on the Dice and ASSD. Furthermore, while the residual block and attention gate are both integrated into E-U-Net, the performances on all metrics improved significantly.

From the boxplot in Fig. 11, we can see that the method's stability gradually improves with the superposition of the model. Compared with the other three networks, the proposed EAR-U-Net has improved on Dice, VOE, and ASSD (Fig. 11 (a) (b) and (d)), and the performance improvement of MSD is the most significant (Fig. 11 (e)). However, multiple outliers caused the proposed EAR-U-Net not to achieve the best performance in RVD. (Fig. 11 (c)).

As for the running time, the network model's training time and testing time increase with the overlay of modules. Nevertheless, such a trade-off way for segmentation accuracy is necessary for clinical application. Fig. 12 shows the loss curves of different models. In the training and verification figures, with the superposition of modules, there is no significant difference between training and verification loss after stabilization, especially the training loss curve almost overlaps.

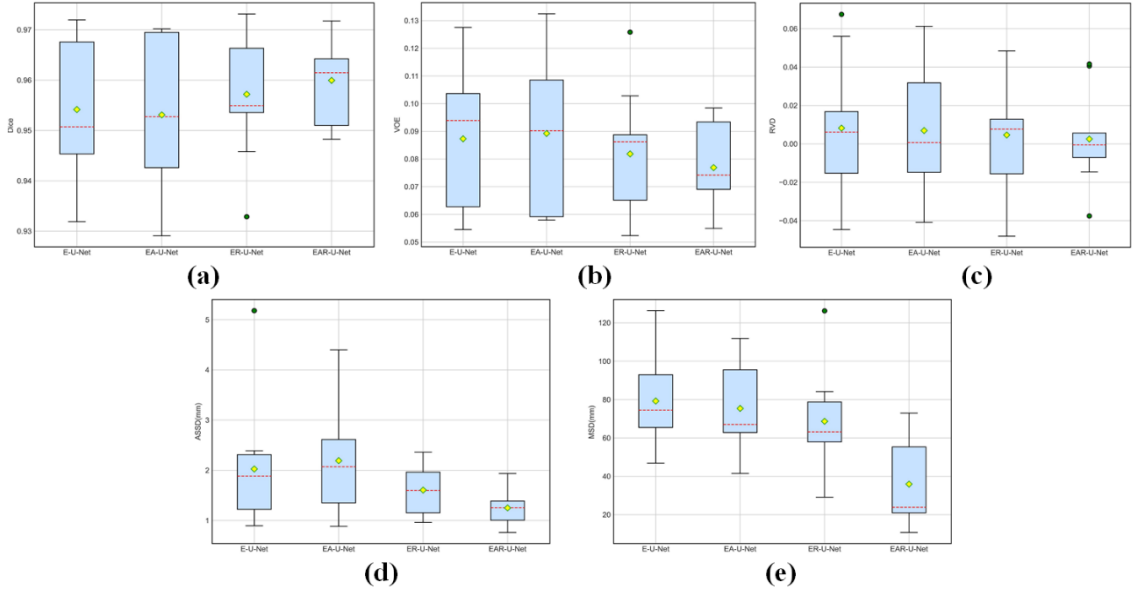

**Fig. 11.** Comparative analysis on evaluation metircs (a) Dice (b)VOE (c) RVD (d) ASSD (e) MSD

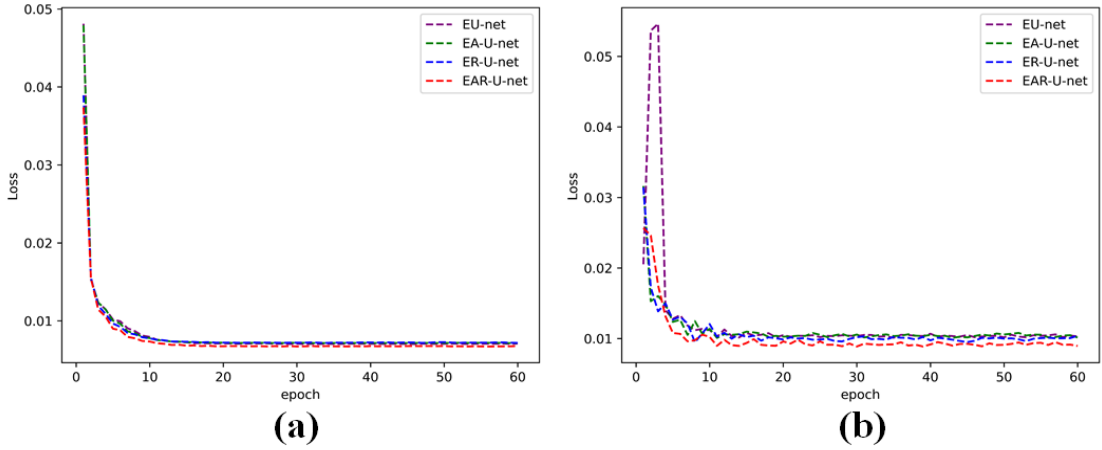

**Fig. 12.** Loss curves of different models in LiTS17 datasets (a) the training set (b) the validation set

### 4.4.3 Evaluation of different loss functions

The loss function is crucial for the training of the model. Both DL and BL perform well in segmentation. In this paper, we assigned DL and BL different weights to train the models in LiTS17-Training datasets. The experimental results listed in Table 7 show that the use of DL performs well on MSD, and the use of BL achieves the best results on RVD. However, given DL and BL a ratio of 1:1, the results show the best performance on Dice, VOE, and ASSD. In terms of training and test time, the impact of loss functions with different weights is slight and negligible. The result analysis of loss functions with different weights is shown in Fig. 13.

**Table 7** Result analysis of different weight loss functions using the EAR-U-Net model

| Loss | Ratio | Dice(%) | VOE(%) | RVD(%) | ASSD(mm) | MSD(mm) | Training time | Test time |
|---|---|---|---|---|---|---|---|---|
| BL | 1 | 96.07±1.06* | 7.55±1.96 | **0.44±2.14** | 1.47±0.67 | 48.28±27.72 | **6h42m48s** | 41.8s |
| DL | 1 | 95.95±0.76* | 7.77±1.42 | 0.5±2.36 | 1.35±0.82 | **35.96±20.62** | 6h45m54s | **41.2s** |
| BL:DL | 0.2:0.8 | 95.84±1.10* | 7.96±2.03 | 1.66±3.08 | 1.67±1.00 | 38.32±14.86 | 6h51m7s | 44.2s |
| BL:DL | 0.5:0.5 | 96.13±0.95* | 7.43±1.75 | 1.29±1.98 | 1.67±0.86 | 43.84±25.54 | 6h48m6s | 43.7s |
| BL:DL | 0.8:0.2 | 96.43±0.90* | 6.88±1.69 | 1.93±2.42 | 1.42±0.72 | 58.03±33.99 | 6h52m45s | 43.6s |
| BL:DL | 1:1 | **96.63±0.82** | **6.50±1.52** | 1.18±2.27 | **1.29±0.35** | 36.79±13.24 | 6h49m55s | 42.3s |

Results are represented as mean and standard deviation. Note: ∗ indicates a statistically significant difference between the marked result and the corresponding one of our method at a significance level of 0.05.

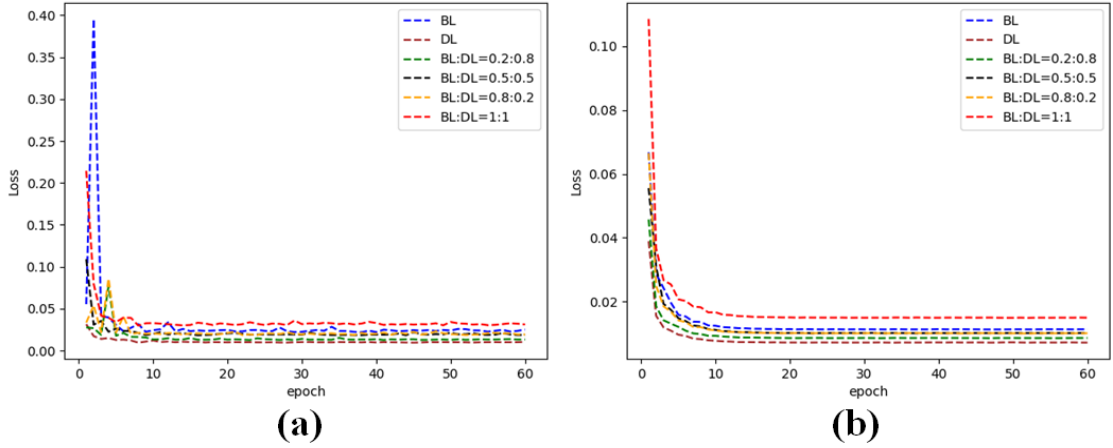

**Fig. 13.** Loss curves of different loss functions on LiTS17 datasets. (a) training set (b) validation set

To verify the segmentation effect of the loss function combined with DL and BL in liver segmentation, we used the weight of DL: BL = 1:1 to test FCN, U-Net, Attention U-Net, and Attention Res-U-Net, respectively, and compared them with DL.

**Table 8** lists the quantitative analysis results of the five models using DL + BL and DL. It can be seen that, compared with the single DL, using DL + BL has improved significantly on Dice, VOE, and ASSD. Specifically, the Dice scores of FCN, U-Net, Attention U-Net, Attention Res-U-Net, and our EAR-U-Net increased by 1.83%, 1.63%, 1.47%, 1.11%, and 0.68%, respectively.

In addition, compared with single DL, using DL: BL = 1:1 enables the standard deviation of all the compared methods on the five evaluation metrics to become smaller. Thus it proves that the DL + BL loss function could improve the segmentation stability. As for training and testing time, the use of different loss functions did not produce significant differences.

Fig.14 shows the loss in the train and validation using DL: BL=1:1 for several classic models. The proposed EAR-U-Net converges the fastest for training loss (Fig. 14(a)), while FCN converges the slowest. For verification loss (Fig. 14(b)), both FCN and U-Net have relatively large volatility in the first few epochs. In contrast, EAR-U-Net has relatively tiny fluctuations, and the loss value is also minimized.

**Table 8** Comparative results of different loss functions with four state-of-the-art methods on 10 LiTS17-Training datasets

| Methods | Loss | Dice (%) | VOE (%) | RVD (%) | ASSD (mm) | MSD (mm) | Training time | Test time |
|---|---|---|---|---|---|---|---|---|
| FCN | DL | 92.46±3.52 | 13.83±5.83 | -1.65±8.74 | 2.86±1.24 | 81.94±28.95 | **4h31m13s** | **33s** |
|  | DL+BL | **94.29±1.9** | **10.75±3.36** | 2.24±5.02 | **2.69±1.16** | **66.36±33.46** | 4h36m22s | 33.4s |
| U-Net | DL | 94.08±2.06 | 11.12±3.65 | -0.48±5.58 | 3.07±2.08 | 66.03±27.91 | 7h49m13s | 36.4s |
|  | DL+BL | **95.71±2.10** | **8.16±3.76** | 3.35±4.22 | **2.06±1.41** | **57.88±29.03** | 7h50m27s | 36.9s |
| Attention U-Net | DL | 94.37±2.27 | 10.58±4.04 | **0.37±6.91** | 2.91±1.57 | 82.03±31.43 | 8h56m35s | 36.8s |
|  | DL+BL | **95.84±1.29** | **7.96±2.38** | 2.45±2.81 | **2.03±1.07** | **71.54±34.06** | 8h58m15s | 36.8s |
| Attention Res-U-Net | DL | 94.93±1.63 | 9.61±2.97 | 2.23±4.12 | 2.77±1.69 | 62.69±19.71 | 9h48m56s | 37.1s |
|  | DL+BL | **96.04±1.03** | **7.60±1.90** | 0.86±3.27 | **1.43±0.47** | **56.99±20.01** | 9h33m7s | 37.7s |
| EAR-U-Net | DL | 95.95±0.76 | 7.77±1.42 | **0.50±2.36** | 1.35±0.82 | **35.96±20.62** | 6h45m54s | **41.2s** |
|  | DL+BL | **96.63±0.82** | **6.50±1.52** | 1.18±2.27 | **1.29±0.35** | 36.79±13.24 | 6h49m55s | 42.3s |

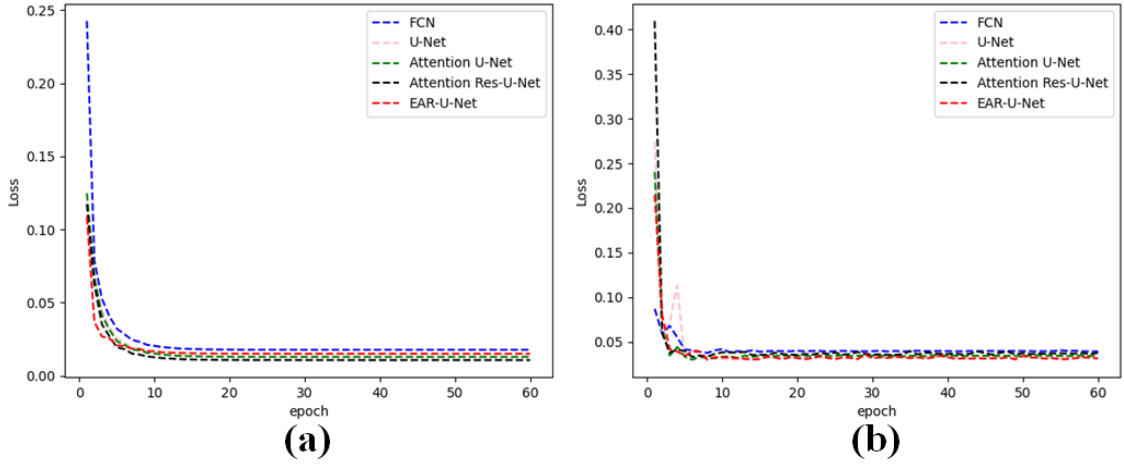

**Fig.14.** Loss curves of two-loss functions on LiTS17 datasets (a) loss in training set (b) loss in the validation set

Fig. 15 shows the visualization of partial segmentation results of FCN, U-Net, Attention U-Net, Attention Res-U-Net, and EAR-U-Net with DL and DL: BL=1:1 as the loss function, respectively.

Fig. 15(a) shows the discontinuous liver region. When DL is used as the loss function, all methods showed over-/under- segmentation errors. In contrast, the errors by all methods are significantly alleviated when DL: BL = 1:1 is used as the loss function.

Fig. 15(b) demonstrates a case of a liver region with adjacent organs of low contrast. We found that the approach using DL as the loss function makes incorrect segmentation at several non-liver organs nearby. However, taking DL: BL = 1:1 as the loss function, only FCN results in noticeable under-segmentation, but the declinations of other models are all greatly improved. Specifically, our proposed EAR-U-Net almost entirely segmented the liver region.

Fig. 15(c) shows a typical case of a small liver region. When taking DL as the loss function, the five methods all showed under-segmentation errors, but the five models almost entirely segment the liver when taking DL: BL = 1:1 as the loss function.

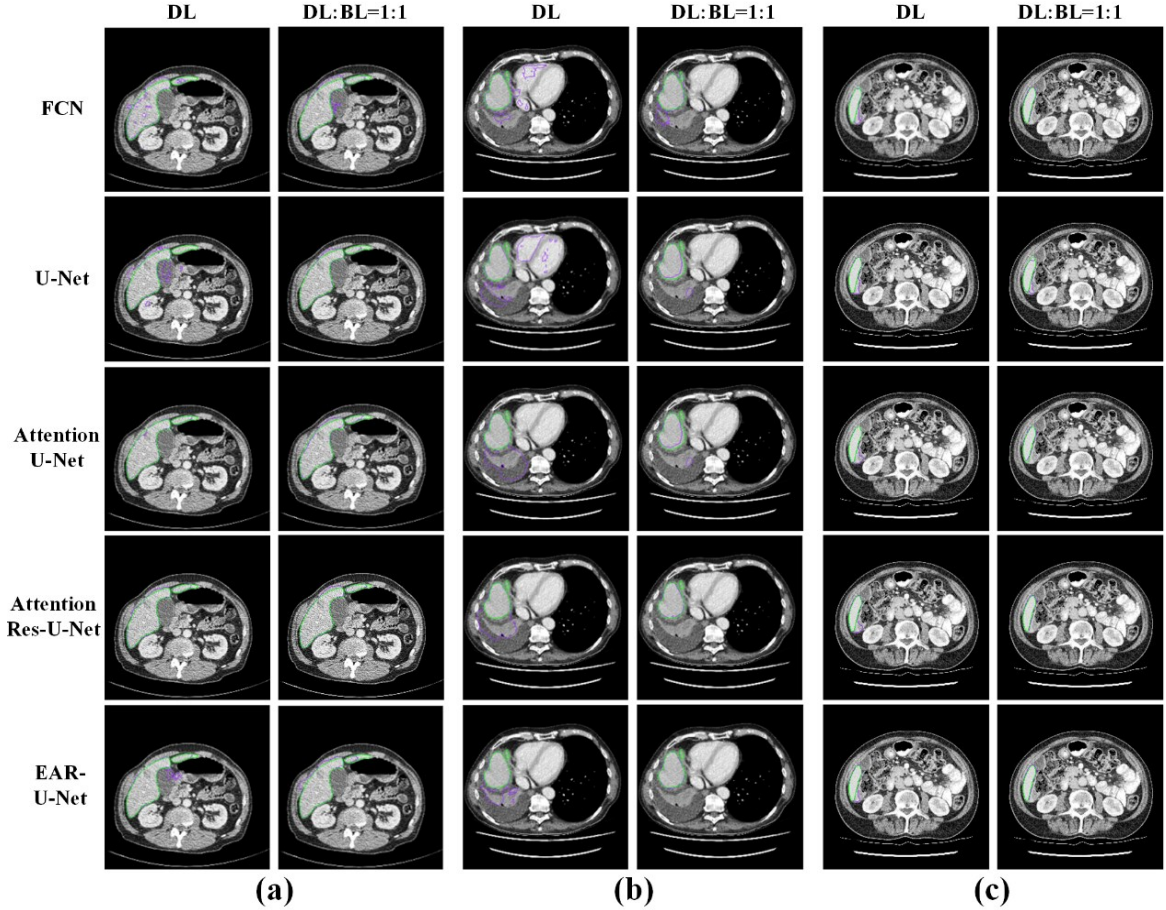

**Fig. 15.** Visualization of typical segmentation cases. (a) discontinuous liver area (b) liver area with the adjacent organs of low contrast (c) small liver area (green line stands for the ground truth, and the purple line represents the result of the corresponding method.)

### 4.4.4 Comparisons of different segmentation methods on LiTS17 test dataset

To further evaluate the performance of the proposed method, we participated in the MICCIA-LiTS17 challenge and compared it with some state-of-the-art methods. The challenge result is shown in Table 9 (Our team's name is hrbustWH402).

As can be seen from Table 9, in the MICCIA-LiTS17 challenge, our proposed method scored 0.952 (ranking 17) and 0.956 (ranking 15) on the two main evaluation metrics of Dice per case (DC) and Dice global (DG), respectively, which is superior to all the listed 2D-based networks. However, our performance is slightly inferior to 2.5D/3D-based networks since our proposed method does not use the 3D inter-slice information.

Table 9 Comparison of various liver segmentation methods in LiTS17 test dataset

| Method | Dimension | DC | DG | VOE | RVD | ASSD | MSD |
|---|---|---|---|---|---|---|---|
| Kaluva et al.[52] | 2D | 0.912 | 0.923 | 0.150 | -0.008 | 6.465 | 45.928 |
| Roth et al.[53] | 2D | 0.940 | 0.950 | 0.100 | -0.050 | 1.890 | 32.710 |
| Wardhana et al.[29] | 2.5D | 0.911 | 0.922 | 1.161 | -0.046 | 3.433 | 50.064 |
| Li et al.[30] | 2.5D | 0.961 | 0.965 | 0.074 | -0.018 | 1.450 | 27.118 |
| Jin et al.[28] | 3D | 0.961 | 0.963 | 0.074 | 0.002 | 1.214 | 26.948 |
| Yuan et al.[54] | 3D | 0.963 | 0.967 | 0.071 | -0.010 | 1.104 | 23.847 |
| **Proposed method** | **2D** | **0.952** | **0.956** | **0.092** | **0.013** | **2.648** | **42.987** |

### 4.5 Test on SLiver07-Training dataset

To verify the generalization capability of the proposed method, we used the weight of DL: BL=1:1 as the loss function and conducted training and testing on the SLiver07-Training dataset. We also compared it with the four classic networks of FCN, U-Net, Attention U-Net, and Attention Res-U-Net. As a result, the proposed EAR-U-Net achieved the best segmentation results in Dice, VOE, RVD, ASSD, and MSD. Specifically, the Dice reached 96.23%. (as shown in Table 10)

Table 10 Quantitative comparison with four state-of-the-art methods on Sliver07-Training datasets

| Methods | Dice (%) | VOE (%) | RVD (%) | ASSD (mm) | MSD (mm) | Training time | Test time |
|---|---|---|---|---|---|---|---|
| FCN | 93.06±1.21* | 12.96±2.11 | -4.47±4.22 | 4.19±2.81 | 114.82±20.58 | **3h22m35s** | **32.5s** |
| U-Net | 95.09±2.83* | 9.01±4.96 | 1.51±3.59 | 1.99±0.87 | 97.62±17.36 | 5h34m49s | 33s |
| Attention U-Net | 95.25±3.14* | 8.94±5.57 | -2.21±3.57 | 2.07±1.63 | 99.85±37.21 | 6h12m23s | 33.4 |
| Attention Res-U-Net | 95.72±2.87* | 8.09±5.11 | -2.06±6.6 | 1.81±0.81 | 103.75±16.56 | 6h58m56s | 33.4s |
| EAR-U-Net | **96.23±2.65** | **7.16±4.75** | **-1.42±5.63** | **1.26±0.68** | **87.32±34.43** | 4h33m2s | 39.5s |

Results are represented as mean and standard deviation. Note: ∗ indicates a statistically significant difference between the marked result and the corresponding one of our method at a significance level of 0.05.

In addition, we also draw a box plot of all the evaluations in Fig. 16, which provides the Dice, VOE, RVD, ASSD, MSD, respectively. The boxplot shows that EAR-U-Net results in the highest median on Dice, and the difference between the upper quartile and the lower quartile is the smallest. For the VOE, we can see that the median of EAR-U-Net is the smallest, while the median of FCN is the largest. For the RVD index, the median of EAR-U-Net is closer to 0. In terms of ASSD and MSD, the lowest median is also obtained by the proposed EAR-U-Net.

Moreover, the proposed EAR-U-NET also shows advantages in network training time. The training time is only 4h33m2s, less than that of U-Net, attention U-Net, and Attention Res-U-Net, but 26% more than FCN. However, the per-case test time is higher than that of other networks.

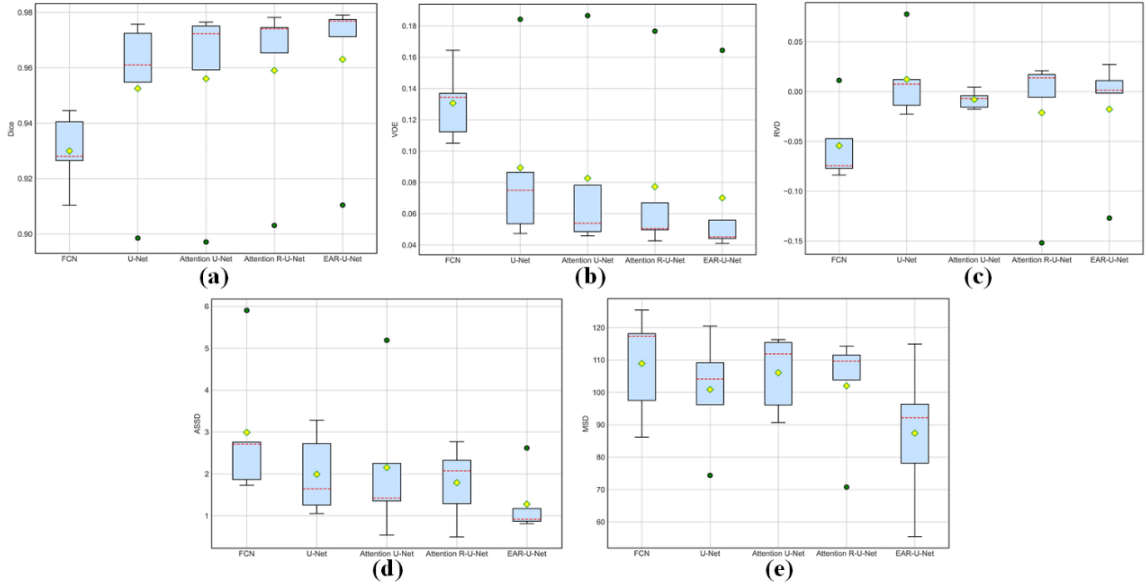

**Fig. 16.** Comparative results of different methods on Sliver07-Training datasets (a) Dice (b)VOE (c) RVD (d) ASSD (e) MSD

Fig. 17 shows the loss curves of training and verification. EAR-U-Net converges the fastest and reduces to the lowest in the training loss. In the loss of verifying set, the loss values of the five networks all show some fluctuations in the first few epochs, but after the loss is stable, the value of EAR-U-Net is reduced to the lowest.

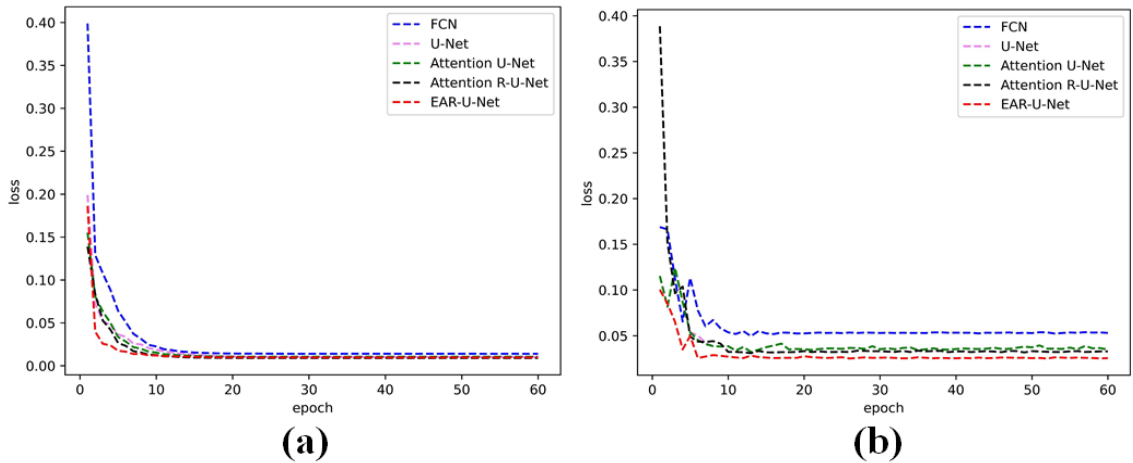

**Fig. 17.** Loss curves of different models on SLiver07 datasets (a) training set (b) validation set.

Fig. 18 shows some visualizations of hard-to-segment livers. (i) The first row is the result of liver segmentation of the gallbladder with similar contrast. It can be seen that FCN, U-Net, and Attention U-Net have mistakenly segmented the gallbladder, while Attention Res-U-Net and EAR-U- Net did not appear to have such an error. (ii) The liver in the second row is adjacent to the low-contrast gallbladder and spleen. It can be seen that FCN segmentation shows the worst effect, not only segmenting the gallbladder but also incorrectly segmenting the spleen far away from the liver. Meanwhile, U-Net also mistakenly segmented the gallbladder. Although the segmentation of attention U-Net and Res-U-Net have improved significantly, there are still some under-segmentation errors. Among all, the segmentation effect of EAR-U-Net is the best. (iii) The third row shows the discontinuous liver area. Again, both FCN and U-Net show obvious over-segmentation errors, while Attention U-Net, Attention Res-U-Net, and EAR-U-Net have alleviated the over-segmentation errors compared with FCN and U-Net. (iv) The fourth and fifth rows demonstrate the liver region containing portal veins. All methods result in specific over-segmentation errors, but the segmentation effect of EAR-U-Net on the portal vein is significantly improved compared to the other four networks. The above cases proved that our network has a better segmentation effect in the liver area containing adjacent organs and portal vein.

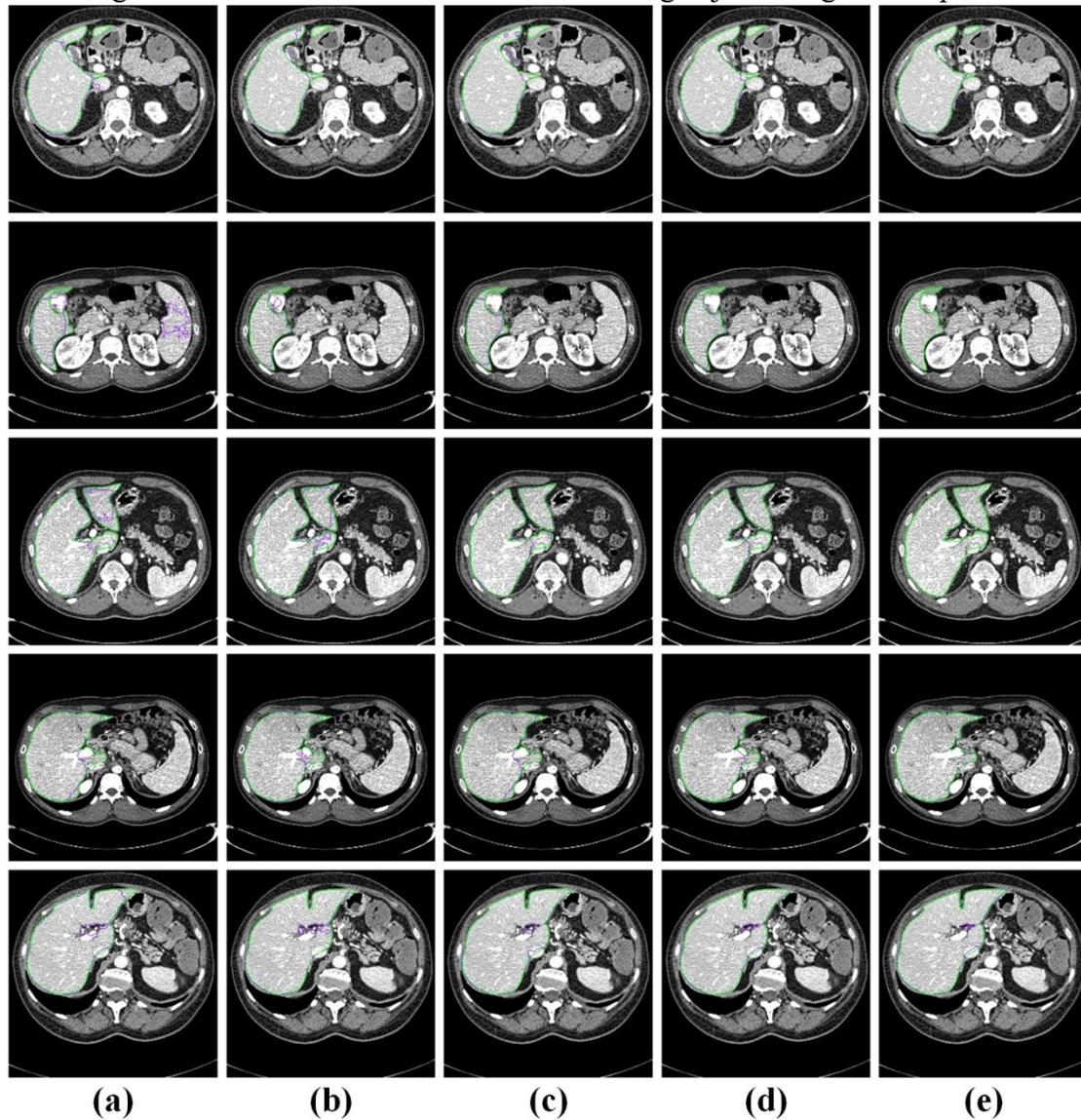

**Fig. 18.** Test on tricky cases of SLiver07. (a) FCN (b) U-Net (c) Attention U-Net (d) Attention Res-U-Net (e) EAR-U-Net (The green line denotes the ground truth, and the purple line indicates the segmentation result of the corresponding method)

### 4.6 Limitations

Although the proposed method achieved satisfactory results, there are still some limitations. As shown in the first column of Fig. 19, there is an apparent inferior vena cava below the liver parenchyma, close to the liver in contrast. Thus the proposed method miss-segment it as the liver region. Furthermore, the second and third columns of Fig. 19 show the presence of liver lesions, which are located at the edge of the liver, and obvious over-segmentation errors occurred by the proposed method. These limitations may be attributed to the inability of the 2D network-based methods to make full use of the 3D inter-splice information. Considering these limitations, we will further conduct 3D-based research in future work and overcome the low-contrast adjacent organs/tumors as the optimization direction.

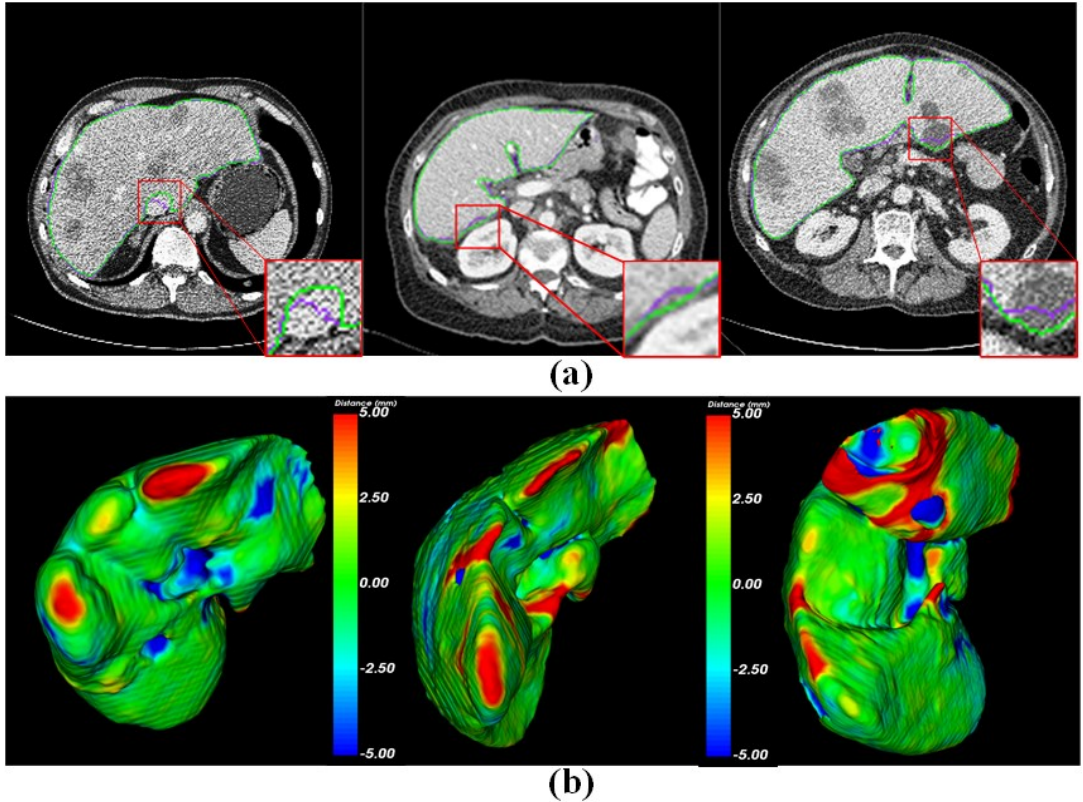

**Fig. 19.** 2D and 3D errors visualizations of the proposed method: (a) 2D errors (green line represents the ground truth, and the purple line represents the segmentation results) (b) 3D errors visualization (blue/red regions indicate an over-/under-segmentation error.)

## 5. Conclusion

This paper presents a new EAR-U-Net network for automatic liver segmentation in CT. To extract feature information more effectively, we employ EfficientNetB4 as the encoder. In addition, to highlight the feature information and eliminate the irrelevant feature responses, we add attention gates to the skip structure. Moreover, the introduction of the residual block also effectively prevents gradient vanishment.

In the experiments, we validated the proposed method on two publicly available datasets, LiTS17 and Sliver07. Specifically, we compared the proposed method with four classical models, including FCN, U-Net, Attention U-Net, and Attention ResU-Net. As a result, the proposed method achieved superior results on five standard metrics. Moreover, we also conducted experiments on different loss functions and proved that the combination of DL and BL produces a better effect in liver segmentation, including challenging cases. However, it is prone to false segmentation in the liver adjacent to other organs/tumors with low contrast.

In conclusion, the proposed EAR-U-Net could enrich the semantic information, enhance feature learning ability, and focus on small-scale liver information. Nevertheless, considering the limitations of the proposed EAR-U-Net in making full use of 3D information, we will focus on the 3D-based segmentation approach for the liver adjacent to organs/tumors with low contrast in future work.